\documentclass[12pt]{article}

\usepackage{amsmath}
\usepackage{amsthm}
\usepackage{amsfonts}

\def\RR{\mbox{\rm I\hskip-.15em R}}

\def\HH{\mbox{\rm I\hskip-.17em H}}
\def\CC{\mbox{\rm C}\hskip-.5em\mbox{l} \;}
\def\id{\mbox{\rm 1\hskip-.28em l}}
\def\gudu{\mbox{$\partial$\hskip-.52em /}}
\def\titel{
\begin{center}
{\Large Mass relations in noncommutative geometry revisited}\\
\large 
Mario Paschke\footnote{e-mail:paschke@dipmza.physik.uni-mainz.de} \\
Institut f\"ur Physik \\
Johannes-Gutenberg Universit\"at \\
55130 Mainz, Germany
\end{center}
}

\begin{document}

\null
\vskip-3cm
\begin{flushright}
\vskip -3cm MZ-TH/97-04 \\
hep-th/yymmddd \\
\end{flushright}

\titel
\vspace{30pt}
\begin{flushleft}
{\bf Abstract:}
We generalize the notion of the 'noncommutative coupling constant' given by
Kastler and Sch\"ucker by dropping the constraint that it commute with the 
Dirac-operator. This leads essentially to the vanishing of the lower bound
for the  Higgs mass and of the upper bound for the W mass. Thus it can be 
concluded that 
these bounds stem from the equal weighting of right- and left-handed fermions.
   \end{flushleft}

\newpage

\section{Indroduction}
One of the interesting features of Connes' version of the standard
model \cite{Connes} is the appearence, at the classical level, of mass 
relations between gauge boson, scalar and fermion masses. In particular one 
obtains a prediction for the Higgs mass as a function of the top mass with 
only a small conceptual uncertainty ('fuzzyness')\cite{KS}.  
However, these results depend on a particular definition of the action, in 
contrast to other predictions (e.g. the occurence of the Higgs field or the 
absence of massive neutrinos), which follow directly from the axioms for 
spectral triples.
For instance the spectral
action defined in \cite{CCh} leads to values for the Higgs mass which differ
from the prediction cited above.\\        
The latter result was obtained from the noncommutative Yang-Mills-action 
${\cal L}_{YM}=(F,F)$, where $(\cdot,\cdot)$ denotes a gauge invariant scalar 
product on the space of two-forms. In this article 
we shall generalize the definition of $(\cdot,\cdot)$ as compared 
to \cite{KS}. Using this generalized scalar product  
one only finds an upper bound for the Higgs mass instead of a prediction.  
We hope that this (more general) point of view leads to a better, physical
understanding of the mass relations, which arise in noncommutative Yang-Mills
theories. 

\section{Noncommutative Yang-Mills theories}
The formulation of Yang-Mills-theories within noncommutative geometry is based
on a spectral triple \cite{Connes} \{${\cal A,H,D},\Gamma,J$\} that is 
constructed as the tensor product of a discrete spectral 
triple \cite{PS1},\cite{Kr} $\{ {\cal A}_{f},{\cal H}_{f},{\cal D}_{f},
\gamma_{f}, J_{f}$\} and the spectral triple 
describing space-time \{$C^{\infty}( {\cal M}),L^{2}({\cal M},S),\gudu_{\cal M},
\gamma_{5},C$\}. Here $L^{2}({\cal M},S)$ denotes the Hilbert-space of 
square-integrable sections of the spin-bundle $S$ and 
$C= i\gamma_{2}\gamma_{0}$ is the operator corresponding to charge conjugation.
Thus we have   \begin{eqnarray*}
{\cal A} &=& C^{\infty}( {\cal M})\otimes {\cal A}_{f}, \\ 
{\cal H} &=& L^{2}({\cal M},S)\otimes {\cal H}_{f},     \\
J &=& C\otimes J_{f},                                    \\      
\Gamma &=& \gamma_{5}\otimes\gamma_{f},                  \\
{\cal D} &=& \gudu_{\cal M}\otimes \id + \gamma_{5}\otimes {\cal D}_{f}.
                   \end{eqnarray*}
\noindent
The representation $\pi({\cal A})$ on ${\cal H}$ induces a representation
of the universal differential algebra $\Omega({\cal A})$ as bounded operators
on ${\cal H}$:
 \begin{eqnarray*}
 \pi: \qquad\qquad    \Omega({\cal A}) &\longrightarrow& {\cal B(H)}\\
          a_{0}{\rm d}a_{1}\cdots {\rm d}a_{n}  & \longmapsto &  
         \pi(a_{0}) \left[ {\cal D}, \pi(a_{1}) \right]\cdots \left[ {\cal D},
                                                       \pi(a_{n})\right]\quad .
\end{eqnarray*}
The differential algebra belonging to the spectral triple, 
$\Omega_{\cal D}({\cal A})$,is then constructed from $\Omega({\cal A})$ by 
dividing out the ideal ${\cal K}$ generated by $ker\:\pi\; 
\cup$\;d$(ker\:\pi)$. 
This raises a technical difficulty since  
$\Omega_{\cal D}({\cal A})$ consists of equivalence classes and therefore it is
quite nontrivial to compute its algebraic structure or, at least, its 
structure as an  ${\cal A}$-bimodule.    
Following \cite{KS} this problem is resolved as follows.\\ 
Being a subspace of ${\cal B(H)}$ , $\pi(\Omega({\cal A}))$ has a natural class 
of bilinear forms :
\[
(A,B)_{z}:= Tr_{\omega}(z A^{*}B |D|^{-4}),          \]
where, a priori, z can be any bounded operator. To avoid technical difficulties 
we restrict $z$ to be of the form $\id \otimes z_{f}$. (In this case one will
recover the classical Yang-Mills-Higgs-action from the action functional 
that will be defined below. We have not examined wether a more general ansatz
for $z$ could also lead to this action.)  $(\cdot,\cdot)_{z}$ will then be 
a nondegenerate
scalar product on $\pi(\Omega({\cal A}))$ if and only if the matrix $z_{f}$ is 
selfadjoint and strictly positive (i.e. $ker\: z_{f}=0$).    
Taking the representative which is orthogonal to ${\cal K}$, it becomes 
possible to identify $\Omega_{\cal D}({\cal A})$ as a subspace of 
$\pi(\Omega({\cal A}))$.\footnote{A sufficient condition for this construction 
to be well-defined is that $\pi({\cal K})$ is closed in the 
closure of $\pi(\Omega({\cal A}))$. It is easy to check, however, 
that there are no serious problems in the case at hand.}
This provides a representation ${\tilde \pi}$ of $\Omega_{\cal D}({\cal A})$ 
as an ${\cal A}-bimodule$ as well as a scalar product $(\cdot,\cdot)_{z}$ on it.
\footnote{ We should mention that, in general, this representation will fail 
to be a representation of the $algebra$ 
$\Omega_{\cal D}({\cal A})$ and even more so of the differential algebra.}\\
Using ${\tilde \pi}$ one can now define the Yang-Mills-action 
                       \[
{\cal L}_{YM} = (F,F)_{z}  
                        \]
where the curvature $F \in {\tilde \pi}(\Omega_{\cal D}^{2}({\cal A}))$ is 
defined as usual $F=$d$A+A^{2}$, and the gauge potential $A$ is a selfadjoint 
element of ${\tilde \pi}(\Omega_{\cal D}^{1}({\cal A})) = \pi(\Omega^{1}
({\cal A}))$. ${\cal L}_{YM}$ will be invariant under the gauge transformations
\[
A \mapsto \pi (u)A\pi (u^{*}) + \pi(u)\pi(du^{*})  \qquad u\in
{\cal U}=\{ u \in {\cal A}: uu^{*}=u^{*}u=1\} 
\]
if the matrix $z_{f}$ commutes with $\pi({\cal A})$. We shall also require 
that it commutes with $\pi^{o}({\cal A}):=J\pi({\cal A})^{*} J^{-1}$.
This makes the parametrization of the matrix $z_{f}$ easier, but it does not 
restrict ${\cal L}_{YM}$: according to the general classification of finite 
spectral triples \cite{PS1}\cite{Kr}
${\cal H}_{f}$ is given as the direct sum of representation spaces for
${\cal A}\otimes{\cal A}^{o}$. Each of these spaces is a tensor product, say
$V\otimes V^{o}$, such that $\pi({\cal A})$ and $\pi(\Omega^{2}({\cal A}))$ 
act as $\id$ on $V^{o}$. $\left[ z_{f}, \pi^{o}({\cal A}) \right]=0$ means then
that the restriction of $z_{f}$ to one of these spaces is of the form 
$\zeta\otimes\id$.\\ 
Recall that the
gauge potential $A$ describes both the vector bosons and the scalar particles
of the theory and that ${\cal L}_{YM}$ gives the complete bosonic part of the
action, possibly including symmetry breaking terms. To define the fermionic 
action one uses the so-called Majorana-spinors, i.e. the elements of the 
subspace:
\[ 
{\cal H}_{real}= \{ \psi \in {\cal H}: J\psi = \psi \}.
\]                  
With the notation $\langle \cdot, \cdot \rangle $ for the scalar product in
${\cal H}$ one sets
\[
{\cal L}_{fermions}:= \langle \psi, ({\cal D}+A+JAJ^{-1})\psi \rangle \qquad
\psi \in {\cal H}_{real}.
\]
${\cal D}_{f}$ can then be interpreted in terms of fermion masses and mixing 
angles which are completely arbitrary unless the choice of ${\cal D}_{f}$
is restricted by an additional principle \cite{PS1}. With our ansatz 
$z=\id \otimes z_{f}$ the coupling constants,
gauge boson masses and scalar masses are determined by the selfadjoint, 
strictly positive matrix $z_{f}$, which has to commute with $\pi({\cal A}_{f})$
and the opposite representation of the algebra $\pi^{o}({\cal A}_{f})$. 
This also implies $\left[ z_{f},\gamma_{f} \right]=0$ as $\gamma_{f}$ is an
element of $\pi({\cal A})\otimes\pi^{o}({\cal A})$. \\
In the series of papers \cite{KS} the authors assumed also 
\begin{equation}
\left[ z_{f}, {\cal D}_{f} \right] = 0.   \label{Zykel}
\end{equation}
This assumption leads to the occurence of lower {\em and}
upper bounds for the W as well as for the Higgs mass. In particular, the two
bounds for the Higgs mass differ only by 34 MeV and one has (for 
$m_{t}=180\pm12$ GeV) the prediction:
\[ m_{H} =288\pm 22 {\rm GeV}.    \]
However, there might be reasons
to drop this condition. For instance, a symmetry requirement \cite{PS1} can 
lead to constaints for $z_{f}$ which are not compatible with (\ref{Zykel}).\\
In the next section we will show that this prediction disappears if one uses 
the generalized scalar product $(\cdot,\cdot)_{z}$, although there is still
an upper bound for $m_{H}$ of about 380 GeV. 

\section{The calculation of the standard model parameters}
The standard model of elementary particle physics is obtained from the 
following spectral data. The {\em real} matrix algebra is chosen as \[
{\cal A}_{f}=\CC\oplus\HH\oplus {\rm M}_{3}(\CC),
 \] 
and is represented on the space ${\cal H}_{f}=\CC^{90}$ with the basis
\begin{eqnarray*}
& & u_{\rm R} \; d_{\rm R}\; u_{\rm L} \; d_{\rm L} \; e_{\rm R}\; e_{\rm L}\;
\nu_{\rm L}\\
& & u_{\rm R}^{c} \; d_{\rm R}^{c}\; u_{\rm L}^{c} \; d_{\rm L}^{c} \; 
e_{\rm R}^{c}\; e_{\rm L}^{c}\; \nu_{\rm L}^{c} \quad.\\
\end{eqnarray*}
Here $u_{\rm R}$, for instance, represents the nine right-handed up-type quarks 
and $u_{\rm R}^{c} =Ju_{\rm R}$ will be identified with their charge conjugate 
by the Majorana-condition $J\psi=\psi$. 
An element $(\lambda,q,m)=a\in{\cal A}_{f}$ acts on these basis elements as
\begin{eqnarray*}
                 & \pi(a): & \\
  u_{\rm R} & \mapsto & \lambda u_{\rm R}\\
  d_{\rm R} & \mapsto & \bar{\lambda} d_{\rm R}\\
  \left( \binom{u_{\rm L}}{d_{\rm L}} \right) 
                     & \mapsto & q\otimes\id_{3}\otimes\id_{N_{f}}
                                  \left( \binom{u_{\rm L}}{d_{\rm L}} \right)\\ 
  e_{\rm R} & \mapsto & \bar{\lambda}  e_{\rm R}\\
  \left( \binom{\nu_{\rm L}}{e_{\rm L}} \right) & \mapsto & q\otimes\id_{N_{f}}
                                \left( \binom{\nu_{\rm L}}{e_{\rm L}} \right)
\end{eqnarray*} 
from the left and as 
\begin{eqnarray*}
                 & \pi^{o}(a): & \\ 
  {\rm Q}    & \mapsto & \left( m^{T}\otimes \id_{3}\otimes\id_{N_{f}} \right)
                                                                   {\rm Q} \\
  \ell & \mapsto & \bar{\lambda}  \ell 
 \end{eqnarray*} 
from the right, where we have used Q for quarks and $\ell$ for 
leptons.  
We refer the reader to \cite{Connes},\cite{KS} or \cite{PS2} for 
further details of this spectral triple. \\
The most general matrix $z_{f}$ leading to a gauge invariant nondegenerate
scalar product is parametrized by ten selfadjoint and strictly positive 
$3\times 3$-matrices $\zeta_{i}$, which act on the different families. Thus 
$z_{f}$ acts on $u_{\rm R}$ by $\id_{3}\otimes\zeta_{1}$, on   
$d_{\rm R}$ by $\id_{3}\otimes\zeta_{2}$, on $\binom{u_{\rm L}}{d_{\rm L}}$
by $\id_{2}\otimes\id_{3}\otimes\zeta_{3}$ and similarly for the other 
particles.\\
Following exactly the lines of \cite{KS} but using the general scalar product
$(\cdot,\cdot)_{z}$, one obtains the following expressions for the coupling
constants :
\begin{eqnarray}
g_{strong} & = & \left[ {\rm tr}(\zeta_{6}+\zeta_{7}+2\zeta_{9}) 
                 \right]^{-\frac{1}{2}} \nonumber \\
g_{weak} & = & \left[ {\rm tr}(3\zeta_{4}+\zeta_{5}) 
                 \right]^{-\frac{1}{2}} \\
g_{hyper} & = & \left[ \frac{1}{2}{\rm tr}(3\zeta_{1}+3\zeta_{2}+\zeta_{3}
                  + \zeta_{8}+2\zeta_{10}) +\frac{1}{6} g_{strong}^{-2}
                 \right]^{-\frac{1}{2}}. \label{gschw}
  \end{eqnarray}
From (\ref{gschw}) and the usual definition $\sin^{2} \theta_{w}=
\frac{g_{weak}^{-2}}{g_{hyper}^{-2}+g_{weak}^{-2}}$
it is immediately clear that we have the bound
\begin{center}
\fbox{\parbox{6cm}{\begin{eqnarray*}
          \sin^{2} \theta_{w} \leq \left[ 1+\frac{1}{6}
                                   \left( \frac{g_{weak}}{g_{strong}}
                                                     \right)^{2}
                                    \right]^{-1}
                    \end{eqnarray*} }}
\end{center}
The matrices $\zeta_{6},\ldots\zeta_{10}$ do not appear in the expressions
for the W and Higgs mass so that $g_{strong}$ and $g_{hyper}$ can be chosen
independently of these two parameters. Before we present these expressions
it is convenient to introduce the following shorthand notations: 
let $m_{i},$ $i=0,\ldots 8$ denote the fermion masses in decreasing order
(i.e. $m_{0}=m_{t}$) , we set 
\begin{eqnarray}
    a_{i}:= \frac{m_{i}^{2}}{m_{W}^{2}}, \qquad i=0,\ldots 8\\
    b_{1}:=\frac{m_{t}^{2}+m_{b}^{2}}{m_{W}^{2}},\quad
    b_{2}:=\frac{m_{c}^{2}+m_{s}^{2}}{m_{W}^{2}},\quad\ldots
    b_{6}:=\frac{m_{e}^{2}}{m_{W}^{2}}.
\end{eqnarray}
For reasons that will become clear later on, we choose the parametrization of
the matrices $\zeta_{1},\ldots,\zeta_{5}$ as follows:
the diagonal elements of $3\zeta_{4}$ and $\zeta_{5}$ are denoted
$\nu_{k},\quad k=1,\ldots 6$, while those of $\zeta_{3}$, $3\zeta_{1}$ and
$3V_{CKM}^{*}\zeta_{2}V_{CKM}$ are denoted by $\mu_{i},\quad i=0,\ldots,8$. 
\footnote{with some suitable ordering of the indices $k,i$}    
The boson masses are then found to be
\begin{eqnarray}
            g_{weak}^{-2} & = & \sum_{k} \nu_{k} \label{schwach}\\
            m_{W}^{2}     & = & \frac{1}{2\sum\limits_{k} \nu_{k}} 
                                   \left[ \sum_{i} \mu_{i} (a_{i}m_{W}^{2}) +
                                        \sum_{k} \nu_{k} (b_{k}m^{2}_{W})
                                      \right] \label{Wma}\\
\nonumber \\
\frac{m_{H}^2}{m_{W}^{2}}& = & 
                \frac{\sum\limits_{i,j} \mu_{i}\mu_{j} (a_{i}-a_{j})^{2}}
                 {\sum\limits_{i,k} \mu_{i}\nu_{k}} +
                 \frac{\sum\limits_{k,l} \nu_{k}\nu_{l} (b_{k}-b_{l})^{2}}
                {2\sum\limits_{l,k} \nu_{l}\nu_{k}} \label{Hima}.
\end{eqnarray}  
Because all the parameters of the right hand side in (\ref{Wma}) are positive 
one gets \begin{center}
 \fbox{\parbox{2.9cm}{$ \frac{m_{e}^{2}}{2}<m_{W}^{2}<\infty.$}}
                           \end{center}  
Now, the problem is to find bounds for the Higgs mass
(\ref{Hima}) under the constraint
\begin{equation}
2\sum_{k}\nu_{k}=\sum_{i}a_{i}\mu_{i} + \sum_{k} b_{k}\nu_{k} \label{Neben}
\end{equation} 
which is just a rephrasing  
of (\ref{Wma}). Choosing $\nu_{6}=1$, $\mu_{0}=\frac{2-b_{6}}{a_{0}}$ and 
all other parameters $a_{i}$, $2-b_{k}$ of the order $\epsilon$ one sees 
from (\ref{Hima}) that the Higgs mass comes out to be of the order $\epsilon$.
Therefore there exists no nontrivial lower bound. To find the lowest 
upper bound under the constraint (\ref{Neben})
is more difficult, but it is straightforward to  prove the following 
estimate.\\
Taking into account the experimental values for the fermion masses one has
the relations   
$(b_{2}-b_{6})^{2}=\min\limits_{k,l >2}
(b_{k}-b_{l})^{2}$ and $\frac{(a_{0}-a_{8})^{2}}{a_{0}+a_{8}}=
\max\limits_{i,j}\frac{(a_{i}-a_{j})^{2}}{a_{i}+a_{j}}$, from which one obtains

\begin{eqnarray*}
\frac{\sum\limits_{i,j} \mu_{i}\mu_{j} (a_{i}-a_{j})^{2}}{\sum\limits_{i,k} 
                                 \nu_{k}\mu_{i}}
& < & (2-b_{6})\frac{16(a_{0}+a_{1})(a_{0}-a_{8})^{2}}{(8(a_{0}+a_{1})+a_{8})
                                (a_{0}+a_{8})} \\ \\
\frac{\sum\limits_{k,l} \nu_{k}\nu_{l} (b_{k}-b_{l})^{2}}{\sum\limits_{l,k} 
           \nu_{k}\nu_{l}}
& < & \frac{(b_{1}-b_{6})^{2}}{2-\eta \frac{4}{5}},\qquad \eta=
               (\frac{b_{2}-b_{6}}{b_{1}-b_{6}})^{2}.
\end{eqnarray*}
As stated above these estimates do not provide 
$\inf\limits_{\mu_{i} \nu_{k}} (m_{H}^{2})$. A suitable choice of the 
parameters $\mu_{i},\nu_{k}$ leads, however, to values of the Higgs mass which 
are very close to the upper bound obtained so far. In a more transparent form 
our result can thus be stated as
\begin{center}
\fbox{\parbox{5cm}{$0<m_{H}^{2}<\frac{17}{4}m_{t}^{2}  
                                     + O(m_{b}^{2}) ,$}}
\end{center} 
where we have only retained the top mass contribution.
\section{Conclusions} 
In \cite{KS} the authors have obtained upper and lower bounds for the 
W-mass as well as for the Higgs mass by using a scalar product on 
$\Omega_{\cal D}({\cal A})$ that was restricted to fulfill the condition
$\left[ z,{\cal D}\right]=0$. We have shown that the upper bound for the 
W-mass and the lower bound for the Higgs mass disappear if one uses a more 
general scalar product.
Now, with our notation, the above condition leads to 
the equations
\begin{eqnarray}
\zeta_{1}=\zeta_{2}=\zeta_{4},\qquad \zeta_{3}=\zeta_{5} \nonumber\\
\zeta_{6}=\zeta_{7}=\zeta_{9},\qquad \zeta_{8}=\zeta_{10} \label{listr}
\end{eqnarray}
which are due to the fact that ${\cal D}_{f}$ maps the right-handed 
fermions to their left-handed partners. 
In addition all the matrices $\zeta_{i}$ have to be chosen diagonal and the 
matrices $\zeta_{2},\zeta_{7}$ have also to be proportional to $\id_{3}$ (if 
one assumes that the CKM-matrix is nondegenerate).
It is clearly the requirement (\ref{listr}) which
leads to the appearence of a {\em lower} bound for the Higgs mass as compared to
our result. We can state that the additional bounds found in \cite{KS}
are a consequence of the equal weighting of particles of different chirality.
The other restrictions on the matrices $\zeta_{i}$ have only a numerical effect
for the bounds we have obtained, but do not lead to a qualitatively different 
result. \indent  
Our last remark concerns the relation for $\sin \theta_{w}$ coming from
(\ref{gschw}). It is possible to generalize the Dirac-operator by taking e.g.
\begin{equation}  
{\cal D}_{\vartheta}:=(\gudu_{\cal M}\otimes \id +\gamma_{5}\otimes 
                       {\cal D}_{f})\cdot(\id\otimes\vartheta),
\end{equation}
if the matrix $\vartheta$ fulfills certain conditions which come from
the axioms for spectral triples. A possible choice of $\vartheta$ would be the
diagonal matrix that multiplies right-handed fermions by $\sin\theta$ and
left-handed fermions by $\cos\theta$ with $\theta\in\RR$. Obviously the
additional parameter $\theta$ would be sufficient to fix $\sin \theta_{w}$
arbitrary. A similar result ( in a different model) was also noted in 
\cite{Andrzej}.\\  
One should mention that there are other deriviations of the
classical standard model Lagrangean from noncommutative geometry, which do not
start from a spectral triple \cite{Rai}\cite{PP}\cite{D-VKM}, see also 
\cite{cM}
for a short review. In particular, Wulkenhaar obtains results which are quite
similar to those of \cite{KS} \\

\vspace*{6pt} \noindent
{\bf Acknowledgements}\\
It is a pleasure to thank F.Scheck for suggesting the subject and decisive help.
We gratefully acknowledge enlightening discussions with W. Kalau.
We would also like to thank A.Sitarz and J.Warzecha for comments and remarks 
on the manuscript.

\end{document}